\documentclass[a4paper,10pt,pre,showpacs,aps,floats,twocolumn]{revtex4}

\usepackage{graphicx,epsfig}

\begin{document}

[Phys. Rev. E {\bf 73}, 036123 (2006)]

\title{The Structure of Peer-to-Peer Social Networks}

\author{Fang Wang $^{1}$, Yamir Moreno $^{2}$, Yaoru Sun $^{3}$}

\affiliation{$^{1}$ Pervasive ICT Research Center, British Telecom, Ipswich
IP5 2TX, UK}

\affiliation{$^{2}$ Institute for Biocomputation and Physics of Complex
Systems (BIFI), University of Zaragoza, Zaragoza 50009, Spain}

\affiliation{$^{3}$ Behavioural and Brain Science Centre, School of
Psychology, University of Birmingham, Birmingham B15 2TT, UK}

\date{\today} 

\begin{abstract}

  This paper presents a statistical analysis of the structure of Peer-to-Peer
  (P2P) social networks that captures social associations of distributed peers
  in resource sharing. Peer social networks appear to be mainly composed of
  pure resource providers that guarantee high resource availability and
  reliability of P2P systems. The major peers that both provide and request
  resources are only a small fraction. The connectivity between peers,
  including undirected, directed (out and in) and weighted connections, is
  scale-free and the social networks of all peers and major peers are small
  world networks.  The analysis also confirms that peer social networks show in
  general disassortative correlations, except that active providers are
  connected between each other and by active requesters. The study presented in
  this paper gives a better understanding of peer relationships in resource
  sharing, which may help a better design of future P2P networks and open the
  path to the study of transport processes on top of real P2P topologies.

\end{abstract}


\maketitle

\section{INTRODUCTION}

In the last several years, many systems have been analyzed unraveling the way
in which their constituents interact which each other. Surprisingly, many
seemingly diverse phenomena found in biological, social and technological
systems \cite{bornholdtbook,vespignanibook,newmanreview,bocareview} share a
complex interaction topology that is in most cases characterized by the
existence of a few key nodes that participates in a large number of
interactions \cite{bornholdtbook,vespignanibook,newmanreview,bocareview}. This
observation is in sharp contrast to previous studies that in order to model the
dynamical aspects of biological, social and technological processes assumed a
regular or a random distribution of interactions for the system's units.
Obviously, the new approach to the topology of networked systems has important
bearings on their dynamics and functioning as have been pointed out during the
last few years \cite{bornholdtbook,vespignanibook,newmanreview,bocareview}. A
first step is then the characterization of the topological properties in order
to get better insights into the dynamics, functioning and new designs of
natural and man-made networked systems.

Peer-to-Peer (P2P) networks form a kind of open, decentralized overlay network
on top of the Internet \cite{vespignanibook}, on which distributed users
communicate directly to find and share resources, often music and movie files.
These networks may be one of the few largest distributed computing systems
ever, and more surprisingly, they can run with great stability and resilient
performance in face of possibly the most ferocious dynamics. The number of
hosts running on Gnutella was reported to be 1,800,000 in August 2005
\cite{Limewire}. Recent studies have extensively investigated the traffic,
shared files, queries and peer properties of some widely applied P2P systems
such as Gnutella and Kazaa \cite{Leibowitz2003, Ripeanu2001, Saroiu2002,
  Zeinalipour-Yazti2002}. It has also been reported that node connectivity (the
number of partners a node interacts with) in Gnutella follows a combination of
a power-law distribution (usually for nodes with more than 10 connections) and
a quasi-constant distribution (for nodes with fewer connections)
\cite{Ripeanu2001}. This may be due to the arbitrarily created connections:
peers establish connections to others by searching presently available peers on
the overlay, in addition to a few links to well known hosts provided by the
system. Peer connections in these systems only suggest routes of traffic and
usually have no relation to peer properties, e.g., peer interests or resources
held by peers.

\begin{table*}
\caption{\label{tab:networkproperties}Topological properties of three (out of six
  studied) original and major peer social networks.}
\begin{ruledtabular}
\begin{tabular}{lllllll}
& SN1 & SN5 & SN6 & SN1 & SN5 & SN6\\
& - original & - original & - original & - major & - major & - major\\
\hline
$N$ & 42186 & 112921 & 191679 & 221 & 459 & 960 \\
$E$ & 81083 & 230500 & 415037 & 666 & 1468 & 3177 \\
\hline
$ \langle k \rangle $   & 3.84  & 4.56  & 4.5   & 6.02  & 6.4 & 6.6\\
$k_{out}$ & 0$\sim$4588 & 0$\sim$7543 & 0$\sim$33680 & 0$\sim$25 & 0$\sim$60 & 0$\sim$71\\
$k_{in}$ & 0$\sim$312 & 0$\sim$765 & 0$\sim$1366 & 0$\sim$50 & 0$\sim$29 & 0$\sim$44 \\
\hline
$ \langle k_w \rangle $ & 8.72 & 10.34 & 11.96 & 13.24  & 17.9 & 15.1\\
$k_{w-out}$ & 0$\sim$28512 & 0$\sim$22510 & 0$\sim$168242 & 0$\sim$244
& 0$\sim$755 & 0$\sim$740\\
$k_{w-in}$ & 0$\sim$7667 & 0$\sim$20326 & 0$\sim$54934 & 0$\sim$114 &
0$\sim$103 & 0$\sim$152\\
\hline
$ \langle w \rangle $ & 2.27 & 2.27 & 2.67 & 2.2 & 2.8 & 2.28\\
$w$ & 1$\sim$1732 & 1$\sim$1719 & 1$\sim$13319 & 1$\sim$102 & 1$\sim$81 & 1$\sim$65\\
\hline
Symmetric links & 12 & 14 & 48 & 12 & 14 & 48\\
\hline
$ \langle l \rangle $   & 5.45  & 6.77  & 8.5   & 3.45  & 4.77  & 6.5 \\
$l_{max}$& 11   & 16    & 19    & 9 & 14 & 17 \\
$ \langle c \rangle $   & 0.019 & 0.021 & 0.015 & 0.09 & 0.092  & 0.091\\
$ \langle b \rangle /N$  & & & & 0.33   & 0.41  & 1.06
\end{tabular}
\end{ruledtabular}
\end{table*}

Recent literature proposed {\em P2P social networks}, to capture social
associations of peers in resource sharing \cite{Wang2005}. Similar to human
social networks, a P2P social network is a collection of connected computing
nodes (peers), each of which is acquainted with some subset of the others. The
social connections of peers indicate that a peer is a resource provider or can
provide information of resource providers to another peer.  Connection
strengths imply the acquaintenanceship or utility of a peer to another, i.e.,
how useful one peer is to another in resource sharing. Although P2P systems
become more and more significant in distributed applications, there is little
knowledge about how peers are socially connected to function together.
Primitive investigation in \cite{Wang2005} confirmed that when peers were
organized according to their social relationships, (instead of arbitrarily
connected links such as those created in Gnutella), the formed P2P networks had
obviously improved search speed and success rate. Moreover, the structure of
P2P social networks is shown to be directed, asymmetric and weighted.

This paper will provide a more comprehensive analysis of peer social networks.
In particular, we report on properties such as degree distribution, clustering
coefficient, average path length, betweenness and degree-degree correlations.
This analysis, on one hand, will give a better understanding of peer
associations in resource sharing and provide hints for future P2P network
design. On the other hand, simulations of transport and other processes
relevant to this kind of network will be enabled from the detailed analysis of
the structure of the networks addressed here.

\section{PEER-TO-PEER SOCIAL NETWORKS}

Several P2P social networks were constructed based on real user information
collected from the Gnutella system. 

An experimental machine running revised Gnucleus, a kind of Gnutella client,
joined the Gnutella network as a super-node, so that it could be connected by
more normal peers and many other super-nodes each of which was also connected
by hundreds of normal peers. In order not to disturb the actual social links
between peers, the experimental node did not provide any shared contents nor
sent queries for resources. It acted as a pure monitor to record the traffic
passing through it. In particular, it recorded information such as which peer
answered a query of which other peer, indicating that the former may be a
useful contact to the latter. The experimental Gnucleus node ran on the
Gnutella network from 5 hours to 3 days. It usually connected 300 normal peers
and 30 other super nodes. The traffic data it recorded involved 1,000 to
200,000 peers.  These data, obviously, only reflected associations of a small
group of peers in the Gnutella system within a limited period of time.  The
Gnutella system should be continuously sampled at multiple points in order to
obtain a more accurate and global picture of peer associations.

The possible social links between peers were discovered from the collected raw
data to form corresponding P2P social networks. A directed connection was
created from peer A to peer B if B was a query answerer of A. The strength or
weight of this connection indicated how many queries B answered A.  The
stronger a connection strength is, the more important the end peer is to the
other peer of the connection. A connection strength with value 1 suggests a
single communication, and hence a weak association. Strength with a constantly
high value suggests the end peer is a frequent resource provider of the start
peer, and hence a long-term and possibly permanent social relation.  The
connection strength, however, may decay over time in the absence of any
contribution from the end peer. This issue was further discussed in
\cite{Wang2005}.

As P2P social networks are directed and the connection strengths indicate peer
affinity, this paper will study P2P social networks in respect of their
undirected, directed (including out and in) and weighted connections. Of
particular interest are the results obtained when the edges are considered
weighted. As most networks in real systems are weighted, it is expected that
their full description will provide a better and more accurate scenario for their
study and modeling. However, the investigation on weighted networks is still a
new area in network modeling, including communication networks, and has only
been addressed recently \cite{barrat04a}.

Table \ref{tab:networkproperties} lists the numbers of nodes (N) and edges (E) of three
out of six P2P social networks studied (marked as SN1 original $\sim$ SN6
original) collected from Gnutella, both at a magnitude of $ 10^5 \sim 10^6$.
The other three are not shown for space reason, but exhibit the same
statistics as of those discussed henceforth. Among tens or hundreds of thousands of
peers, only a few of them acted as both requesters and providers. These peers
play a major role in P2P social networks as they contribute essential links
to the networks. These peers are hence called major peers. Table
\ref{tab:networkproperties} also shows the information of the social networks of major peers
(marked as SN1 major $\sim$ SN6 major), refined from the above original social
networks, respectively.  The number of major nodes and their edges is only of
$10^2 \sim 10^3$. For instance, the number of nodes in the major network
obtained from SN1 drops from 42,186 to only 221. In the remaining of this
paper, both original P2P social networks and major peers' social networks will
be investigated.

\section{STATISTICAL ANALYSIS}

\subsection{Connectivity properties}

Table \ref{tab:networkproperties} gives a summarization of the average degree $
\langle k \rangle $, range of out degrees $k_{out}$ and in degrees
$k_{in}$ for the unweighted representations of P2P networks
analyzed. In the case of weighted representations, the table shows the average
weighted degree or strength $ \langle k_w \rangle =
\sum_j w_{ij}+\sum_j w_{ji} $ and range of weighted out $k_{w-out}$ (the
first term in the sum) and in $k_{w-in}$ degrees (the second term in
the sum) of the original and major P2P social networks studied. Here,
$w_{ij}$ is the weight of the $ij$ link and means that $j$ answered
$w_{ij}$ queries from $i$. The average connection weight $ \langle w
\rangle $, the weight range $w$ and the number of symmetric links are
also listed in this table.

\begin{table*}
  \caption{\label{tab:peerpercentage}Percentage of peers with null, 1, 2 and
    more out and in degrees. Note that there are much more resource providers
    than requesters.}
\begin{ruledtabular}
\begin{tabular}{lllll}
$k =$   & 0 & 1 & 2 & $>$2 \\
\hline
Out (original)  & 98.5$\pm$0.02\%   & 0.16$\pm$0.04\%   & 0.07$\pm$0.001\%  & $<$1.27\% \\
In (original)   & 0.86$\pm$0.03\%   & 68.5$\pm$4.3\%    & 14.6$\pm$1.7\%    & $<$16.1\% \\
\hline
Out (major) & 42$\pm$2.6\%  & 17.7$\pm$1.6\%    & 8.6$\pm$1.1\% & $<$31.7\% \\
In (major)  & 15$\pm$2.5\%  & 33$\pm$2.4\%  & 15.2$\pm$1.2\%    & $<$36.8\% \\
\end{tabular}
\end{ruledtabular}
\end{table*}

Each peer in the original peer social networks has an average of 4.3$\pm$0.22
neighbors.  This also means that on average a peer has 2.15 out degrees and in
degrees.  This number slightly increases with the number of peers, but is very
small compared with a fully connected network of the same size $ \langle k
\rangle =N-1 \sim 10^{5\sim6}$. Some peers, however, have up to nearly three
thousands to tens of thousands out connections (i.e., resource providers),
while the maximum connected resource requesters (i.e., in degree) of a peer is
only hundreds up to one thousand. This suggests that there are generally more
available providers, though a provider only serves a small fraction of peers in
the network. The average weighted degree is around 9$\sim$12 per node and the
average connection weight is around 2.3. That is, the frequency of a peer to
contact another is about 2.3 times, though in reality a peer can answer another
peer's requests as many as ten thousands times.

\begin{figure}
\epsfig{file=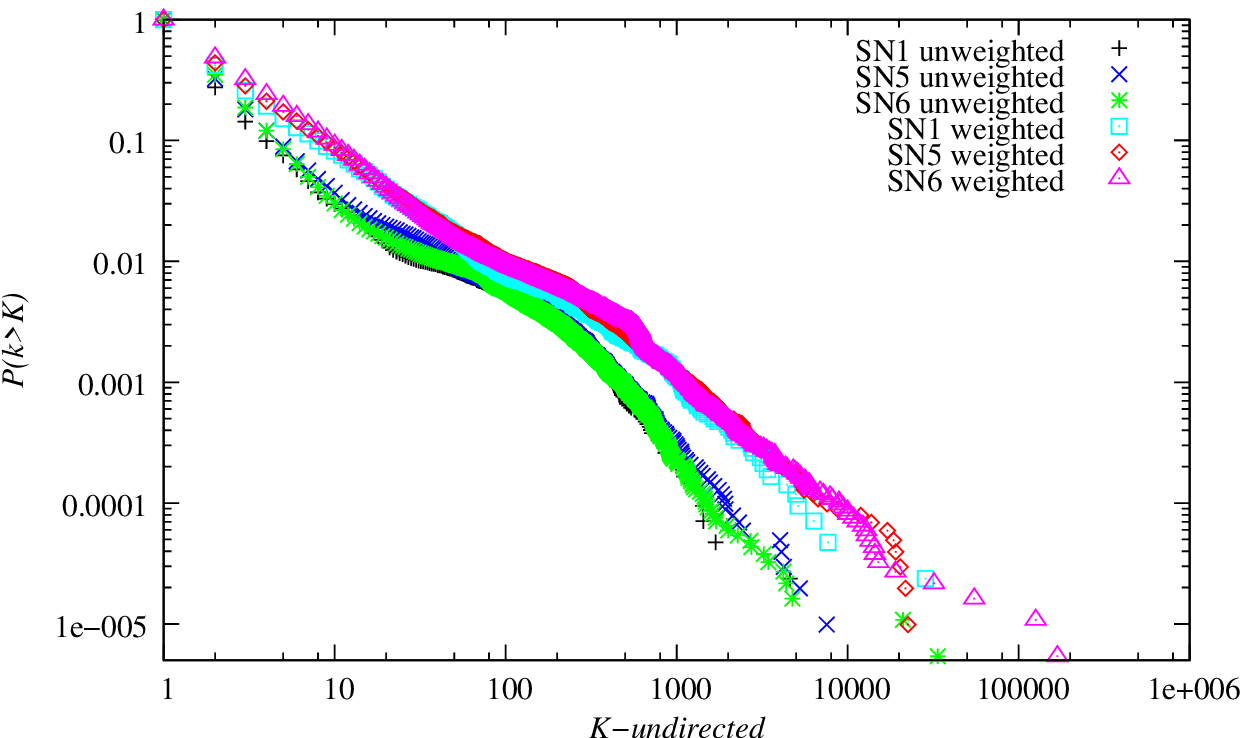,width=3.3in,angle=0,clip=1}
\epsfig{file=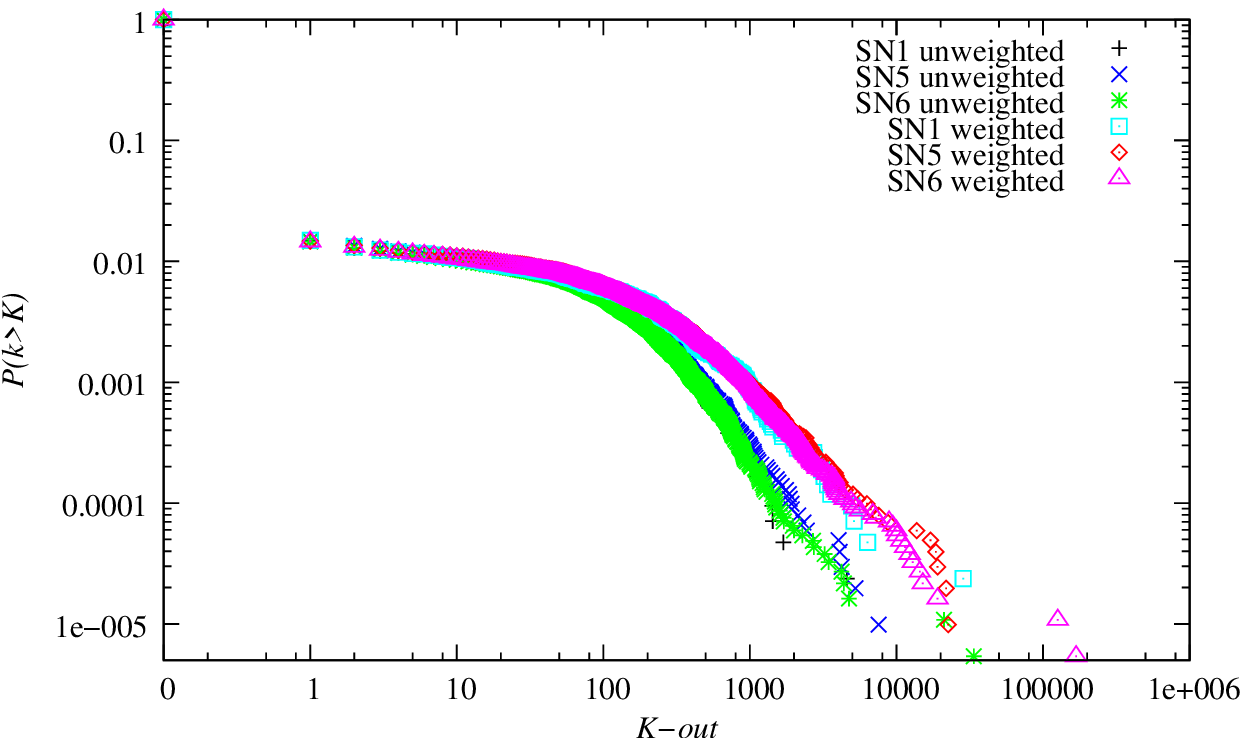,width=3.3in,angle=0,clip=1}
\epsfig{file=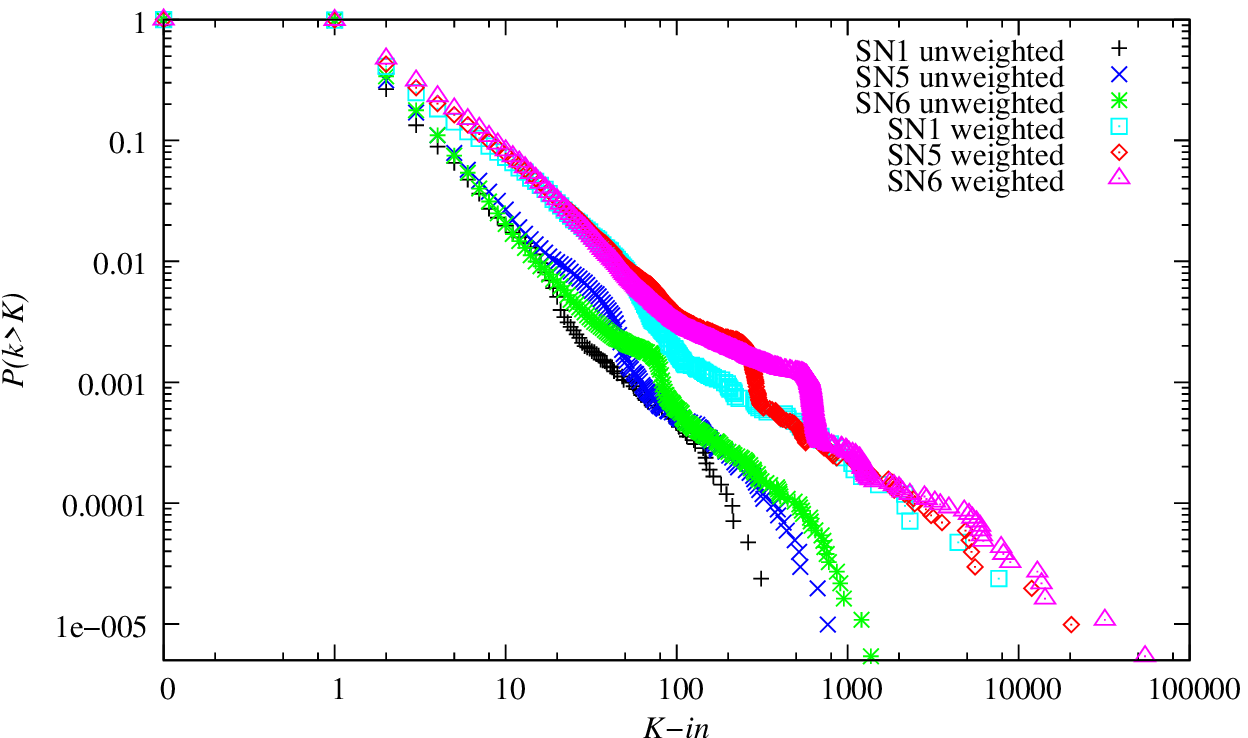,width=3.3in,angle=0,clip=1}
\caption{ Cumulative undirected, out and in degree distributions for three P2P
  networks and their weighted representations. Values of the exponents
  characterizing the (power-law) distributions are reported in Table
  \ref{tab:exponents}. Note that although SN1, SN5 and SN6 are different networks,
  they all fall in what seems to be a universal curve.}
\label{fig:degreedistribution}
\end{figure}

Similar results have been shown in major peers' social networks. The social
networks of major peers are denser than the original ones, as the average
connectivity is almost doubled among major peers. The average connection
strength of major social networks is nearly the same as that of the original
social networks, suggesting that the average level of peer acquaintance is
independent from network sizes. While there are hundreds of connections present
in the network, only few of them have symmetric links, less than 0.03\% of the
whole connections and all the symmetric connections are between major peers.
This proves that real peer social networks are extremely asymmetric: while one
peer presents a useful social contact to another, it is seldom the case in
which the other deems that one as its useful supplier.

Table \ref{tab:peerpercentage} lists the percentage of peers that have no or 1,
2 and more out and in connections in both original and major social networks.
Significantly, 98.5\% of peers have no out neighbors at all. These peers are
pure providers that never requested anything. Accordingly there are only 0.86\%
peers that did not answer any request of others. 68.5\% of the peers answered
one query and more than 30\% peers answered more. A similar phenomenon has also
been found in major peers networks. The above result, namely, the fact that
there are much more resource providers than requesters, points to an important
structural property that may be at the root of the high reliability of Gnutella
despite the system's extreme dynamics and uncertainty.

The degree distributions of undirected, out and in connections have also been
investigated.  Fig. \ref{fig:degreedistribution} illustrates unweighted and
weighted degree distributions of the original social networks SN1, SN5 and SN6
respectively.  (Social networks of major peers present very similar degree
distributions so they are not shown here due to the lack of space.) The results
confirm that peer social networks follow power-law distributions and the
exponents are summarized in Table \ref{tab:exponents}.

It is worth noting that a universal exponent has been obtained for each group
of networks (see Fig. \ref{fig:degreedistribution}), namely P2P social networks
show the same exponent of the degree distribution for undirected connections no
matter of their specific characteristics (e.g., size, number of edges, etc) and
the same holds for directed and weighted distributions. Moreover, weighted
networks exhibit similar degree distributions, though statistically different
as far as the exponent of the power law distribution is concerned, to those of
unweighted networks.  For six peer social networks and corresponding major
networks, their out degree distributions have an average exponent of $\gamma
\approx 1<2$, and both in and undirected degree distributions have an exponent
$\gamma>2$. This is an interesting feature as $\gamma=2$ forms a dividing line
between networks with two different dominating behaviors. Hence the different
power-law distributions obtained here suggest that the average properties of
peer social networks are dominated by (requesting) individuals that have a
large number of providers, while providing peers with fewer connected
requesters dominate the provision flow of resources.

Recent studies reported that the underlying peer-to-peer Gnutella network has
degree exponent less than 2 \cite{Allaei2005}\cite{Jovanovic2001}, contrary to
the undirected degree exponent of P2P social networks found in our work.  While
global information exchange mechanisms are closely related to networks with
exponent $\gamma<2$ \cite{Allaei2005}, P2P social networks may involve more
local interactions between associated peers. However, peer social networks
won't prevent global interaction and information diffusion (e.g., web caches)
if required. It would be interesting to see the performance and topological
changes when P2P social networks are incorporated with those global mechanisms.

\begin{table*}
  \caption{\label{tab:exponents}Exponents $\gamma$ for
    undirected, directed and weighted representations of P2P social networks.}
\begin{ruledtabular}
\begin{tabular}{llll}
$\gamma$            & Undirected    & Out       & In \\
\hline
Original unweighted & 2.1$\pm$0.07  & 0.95$\pm$0.12 & 2.6$\pm$0.13 \\
Major unweighted    & 2.53$\pm$0.096    & 1.14$\pm$0.18 & 2.65$\pm$0.062 \\
Original weighted   & 2.98$\pm$0.026    & 0.92$\pm$0.09 & 2.2$\pm$0.11 \\
Major weighted      & 2.13$\pm$0.1  & 1.03$\pm$0.14 & 2.2$\pm$0.14 \\
\end{tabular}
\end{ruledtabular}
\end{table*}

\subsection{Average shortest path lengths and betweenness}

The shortest distances between all pair of peers that have (directed) paths
from one to another are calculated. The average distances of the shortest paths
in the original and major social networks are around 6.6 and 4.6 respectively,
as shown in Table \ref{tab:networkproperties}. Here the law of six degrees of
separation still come into existence in spite of the huge sizes and sparseness
of the peer social networks.  The social networks of major peers are obviously
better connected. In general, a major peer can reach another randomly chosen
major peer in around 4.6 steps. The smaller average shortest path length of
major peers is of the order one may expect from the logarithmic dependency of
$\langle l \rangle $ with $N$ in small-world networks. Another possible
explanation is that major peers show disassortative correlations. This kind of
correlations happens when nodes of different degrees are likely connected.
That is, there is no core that concentrates all major peers. Otherwise, one
would expect a greater decrease in the average shortest path lengths than that
observed.  This hypothesis will be confirmed in the following analysis on
degree-degree correlations, which shows that, within statistical fluctuations,
peer social networks are mainly disassortative.

The average path lengths of both original peer social networks and major peer
social networks are much smaller than those for a regular two-dimensional
lattice of the same size, which range from tens to hundreds. It has been found
that the average distances vary logarithmically with the number of individuals
in some kinds of social networks including scientific collaboration networks
\cite{Newman2001, Watts1998}. Unfortunately, our data are too sparse to confirm
or reject this. (However, as shown in the tables, $ \langle l \rangle $ is
certainly small in all cases.) Analysis of more peer social networks may be
helpful.

\begin{figure}
\epsfig{file=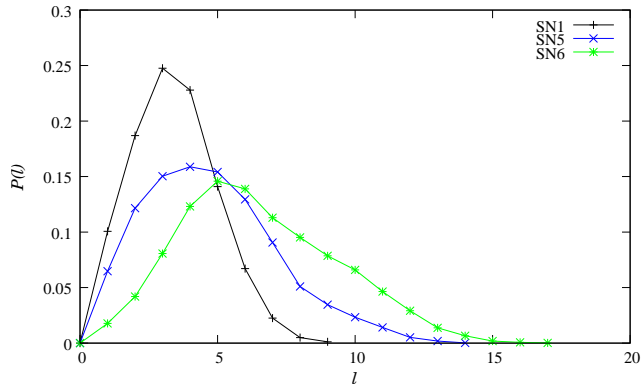,width=3.3in,angle=0,clip=1}
\caption{ Frequency of average shortest path lengths in major peer social
  networks.}
\label{fig:shortestpath}
\end{figure}

The maximum distance $l_{max}$ between connected peers, or diameter of the
network, is on average 14.5 for original social networks and 12.5 for major
peer networks. This suggests that connected peers in these networks can be
reached by a chain of at most 15 or 13 acquaintances. Fig.
\ref{fig:shortestpath} illustrates the frequency of the shortest paths in
social networks SN1, SN5 and SN6 respectively. These shortest paths have a long
tail, which distinguishes peer social networks from random networks with the
same number of nodes and edges. The long tail of the shortest path has been
reported as a property of small world networks \cite{Adamic1999}.

A property closely related to the distribution of average shortest path lengths
is the betweenness. The betweenness measures the centrality of a node in a
network and allows exploration of the influence a node has over the spread of
information through the network. It is normally calculated as the fraction of
shortest paths between node pairs that pass through the node of interest.
Betweenness is commonly applied in social network analysis, and has been
recently introduced for load analysis in scale-free networks \cite{Goh2001}. A
direct calculation of peer betweenness in the original peer networks is rather
laborious due to the enormous number of peers involved. Here only the average
betweenness $ \langle b \rangle /N$ of the major peers social networks is
presented in this section, as listed in Table \ref{tab:networkproperties}. The
average betweenness over major peers is between $0.3N \sim N$, indicating that
the social networks are not dominated by a few highly connected peers. 

\begin{figure}
\epsfig{file=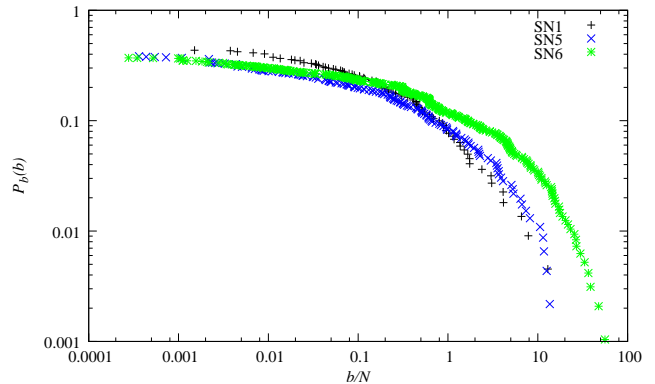,width=3.3in,angle=0,clip=1}
\caption{Cumulative betweenness distribution of the undirected representation
  of three major P2P networks.}
\label{fig:betweenness}
\end{figure}

\begin{figure}
\epsfig{file=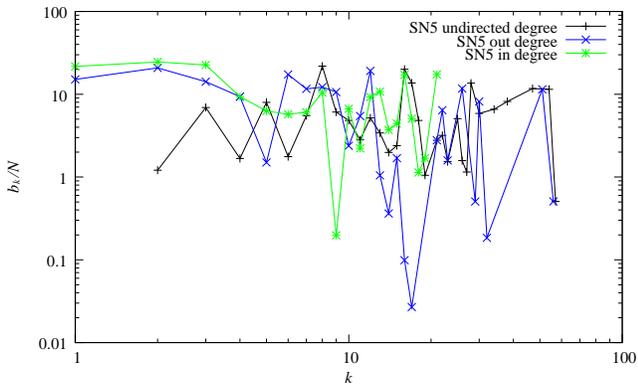,width=3.3in,angle=0,clip=1}
\caption{Betweenness $b_k$ as a function of the peer's connectivity
  $k$. Note the lack of any scaling of $b_k$ with $k$. See the text
  for further details.}
\label{fig:1214bk}
\end{figure}

We further investigated betweenness distribution $p(b)$, the probability that
any given peer is passed over by $b$ shortest paths (see Fig.
\ref{fig:betweenness}) and the relationship between the average betweenness of
a peer and its connectivity $k$ (see Fig.  \ref{fig:1214bk}).  Again, no clear
power-law decay for the former or a linear increase for the latter has been
found, as previously reported for other networks \cite{Goh2001, Vazquez2002}.
In our case, the fact that $b_k$ does not scale with $k$, and hence, the lack
of any correlations important for information traffic and delivery, is another
indication of the unique topological properties of these networks, making their
functioning very reliable and robust. It is worth noting at this point
  that an interesting and relevant issue to be explored more carefully in
  future works is whether or not self-averaging verifies in these systems.
  While Figs. \ref{fig:shortestpath} and \ref{fig:betweenness} may suggest the
  lack of self-averaging, they correspond to major networks, which are still 
  too small to draw definitive conclusions. Moreover, the intrinsic dynamic
  nature of these networks may perfectly reconcile networks properties that are
  not sample-dependent (e.g global properties such as degree distributions)
  with other local metrics that depend on the sampling (as those depicted in
  Figs. \ref{fig:shortestpath} and \ref{fig:betweenness}).

\subsection{Clustering coefficient}

The clustering coefficient is an important local network property that measures
how well the neighbors of any node in a network are locally connected. Table
\ref{tab:networkproperties} gives the values of clustering coefficients of the
networks studied here. Original peer social networks possess a similar
clustering coefficient $ \langle c \rangle \approx 0.02$.  This small number
suggests that peer neighbors are not closely connected, i.e., only a few
neighbors deem others as their acquaintances. However, the closeness of peer
social networks is better than ER random graphs with the same size and average
connectivity, whose clustering coefficients are $ \langle c \rangle _{rand} =
\langle k \rangle /N \approx 10^{-5}$, three orders of magnitude less than
those of the peer social networks. 
At the same time, the estimate for the
  clustering coefficient might be consistent with that of random graphs with
  scale-free degree distribution. 
Compared with the original social networks,
major peers show closer relationships with each other.  The clustering
coefficients of major peers are nearly 0.1, one to two magnitudes larger than
their corresponding random graphs. Thus the active players in peer social
networks, which both provide and request resources, are themselves relatively
well connected.

The clustering coefficients are kept constant for peer social networks or major
peers social networks with different sizes, suggesting there may be a unique
value to them, a property that has been observed in other systems as well
\cite{bornholdtbook,newmanreview}. Moreover, the highly clustered property and
short paths between distributed peers (as introduced in Section III.B) confirm
that peer social networks are small worlds, as other natural or artificial
networks, such as ecosystems, human societies and the Internet
\cite{bornholdtbook,vespignanibook,newmanreview,bocareview}.

\begin{figure}
\epsfig{file=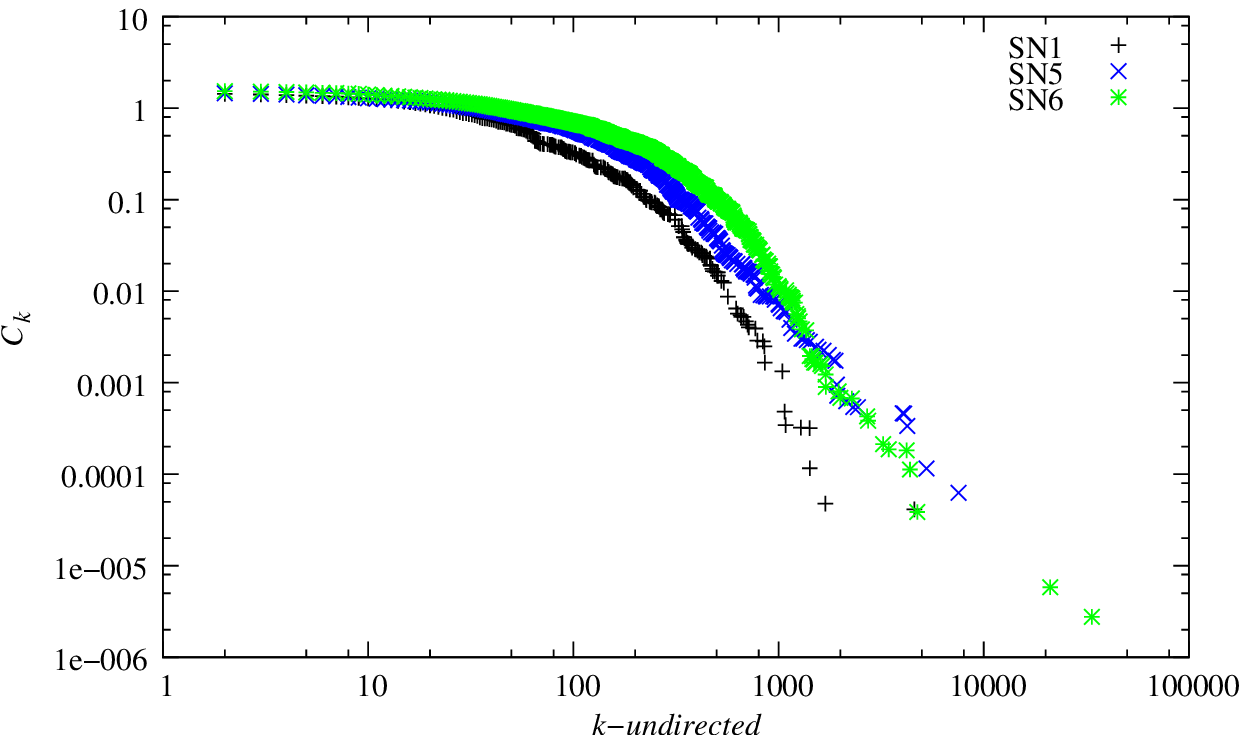,width=3.3in,angle=0,clip=1}
\epsfig{file=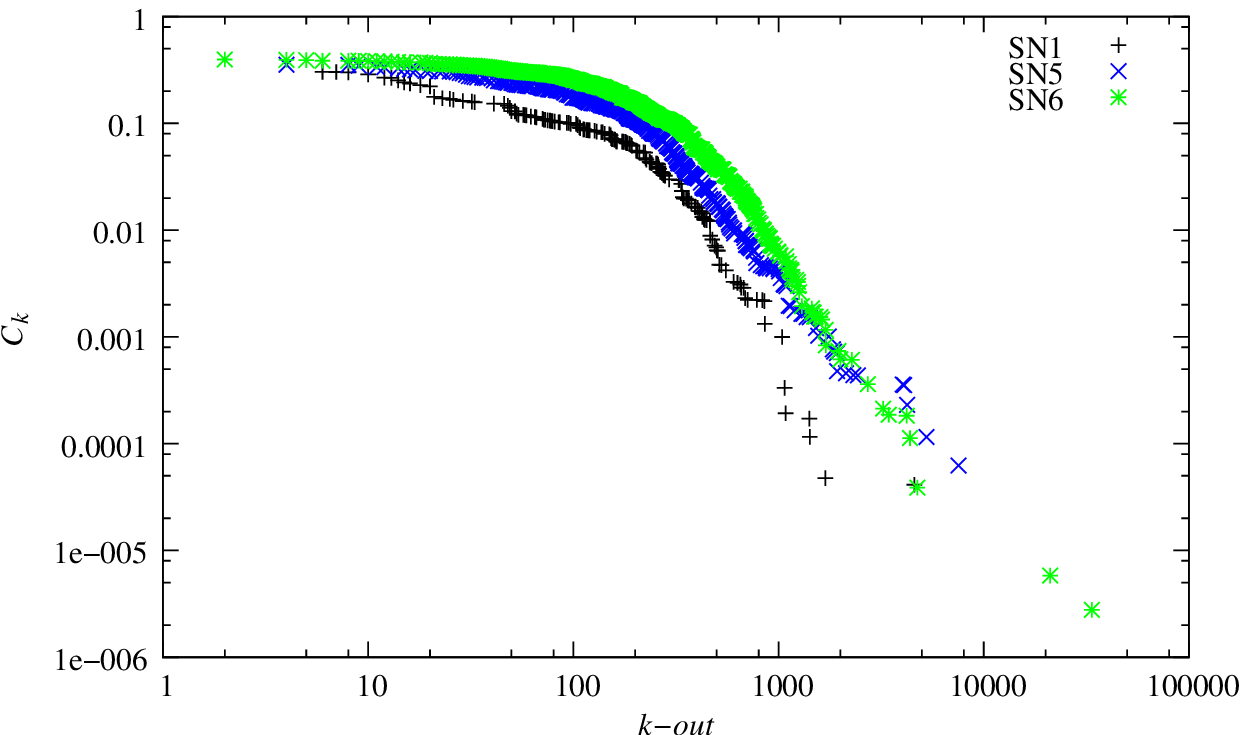,width=3.3in,angle=0,clip=1}
\epsfig{file=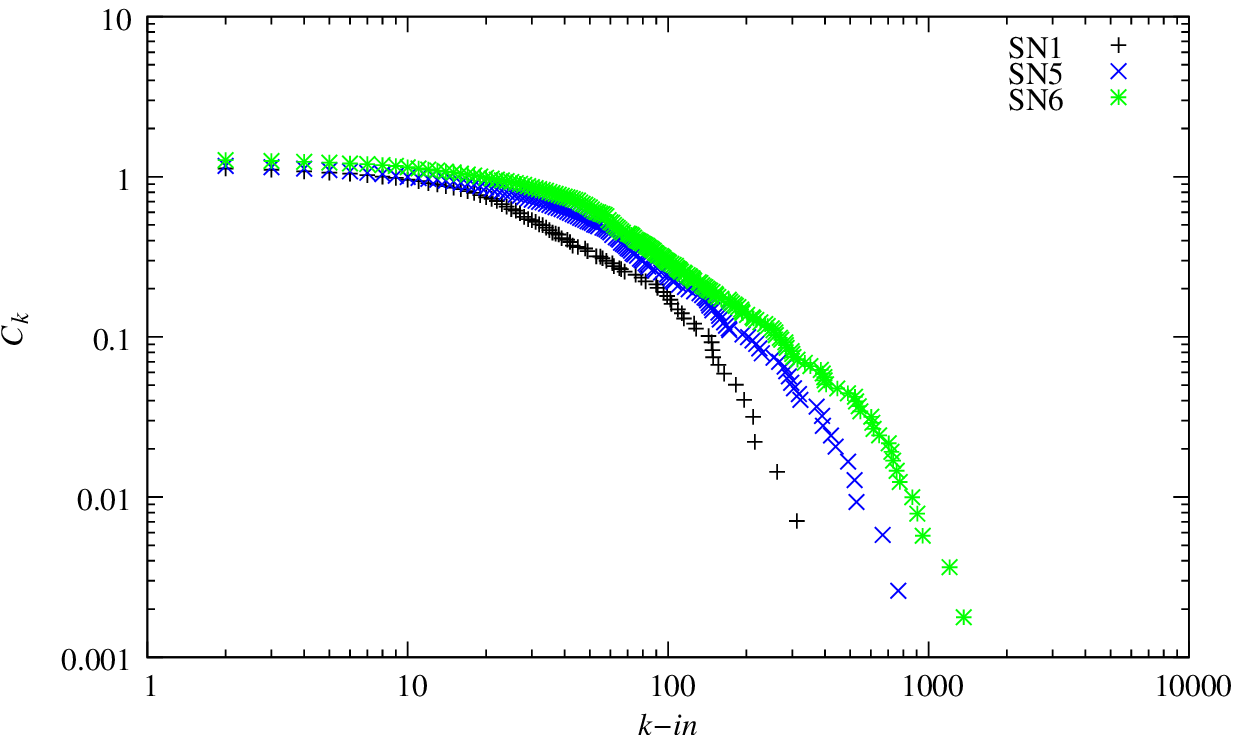,width=3.3in,angle=0,clip=1}
\caption{ Cumulative clustering coefficient $C_k$ as a function of
undirected, out and in degrees $k$.}
\label{fig:cumucc}
\end{figure}

Studies on scientific collaboration networks and Internet topologies reported a
power-law relationship between the average clustering coefficient $C_k$ over
nodes of degree $k$, that is, $C_k \sim k^{-a}$ \cite{Newman2001, Vazquez2002}.
Fig. \ref{fig:cumucc} plots $C_k$ of some original peer social networks in
relation to peers' undirected, out and in degrees. A clear power-law form is
difficult to claim in our data. Nevertheless, the non-flat clustering
coefficient distributions shown in the figures suggest that the dependency of
$C$ on $k$ is nontrivial, and thus points to some degree of hierarchy in the
networks. Further study of social networks' hierarchy will clarify this point
and will be undertaken in future work.

\subsection{Degree-degree correlations}

Networks with assortative mixing are those in which nodes with many connections
tend to be connected to other nodes with many connections and vice versa.
Technological and biological networks are in general disassortative, and social
networks are often assortatively mixed, as suggested by the study on scientific
collaboration social networks \cite{Newman2001}. Contrasting to this however,
Internet dating communities, a kind of social network embedded in a
  technological one, displayed a significant disassortative mixing
\cite{Holme2004}. This seems to be our case as well.

Table \ref{tab:mixingcoefficients} lists the correlation coefficients of all
types of degree-degree correlations for both original peer social networks and
networks of major peers. Correlations are measured by calculating the Pearson's
correlation coefficient $r$ for the degrees at either side of an edge:

\begin{equation}
r=\frac{ \langle k_{out}k_{in} \rangle - \langle k_{out} \rangle
\langle k_{in} \rangle }{\sqrt{ \langle k^2_{out} \rangle - \langle
k_{out} \rangle ^2}\sqrt{ \langle k^2_{in} \rangle - \langle k_{in}
\rangle ^2}}
\end{equation}

Similar to Internet dating communities, peer social networks present
dissortative mixing when directions are not considered in peer
connections.  Positive mixing is shown for $r_{inout}$ and $r_{inin}$
in most social networks, suggesting that active requesters (with a
high $k_{out}$) tend to associate active providers (with a high
$k_{in}$), and even active providers tend to associate with each
other. Between major peers that both provide and request resources,
active requesters also have a preference towards each other. It is not
surprising that $r_{outin}$ is always negative in both original and
major peer networks, which means that providers with many requesters
are actually less often associated with frequent requesters. The
generally dissortative mixing property of peer social networks
suggests that peer networks in general are vulnerable to targeted
attacks on highest degree peers but a few attacks on some providers
may not destroy the network connectivity due to the existence of other
providers in the core group.

\begin{table*}
  \caption{\label{tab:mixingcoefficients} Correlation coefficients for original
  and major peer social networks. Negative figures indicate that
  poorly connected nodes are likely linked to highly connected nodes
  while positive values mean that connectivity peers tend to connect
  to each other.}
\begin{ruledtabular}
\begin{tabular}{lllllll}
& SN1 & SN5 & SN6& SN1 & SN5 & SN6 \\
& - original &  - original & - original& - major  & - major & - major\\
\hline
$r$& -0.091&  -0.095& -0.109  & -0.018& 0.014& -0.048\\
$r_{inin}$& 0.028 & 0.014& 0.028    & 0.019 & 0.126 & -0.004 \\
$r_{inout}$& 0.007& 0.003& 0.008 & -0.016& -0.006& 0.019\\
$r_{outin}$& -0.098& -0.102& -0.106& -0.074& -0.088& -0.106  \\
$r_{outout}$& -0.023& -0.01& -0.025& 0.052& 0.09& 0.054\\
\end{tabular}
\end{ruledtabular}
\end{table*}

\section{Conclusions and future work}

This paper presents the first study on social associations of
distributed peers in Peer-to-Peer networks. Several peer social
networks have been constructed from the real user data collected from
the Gnutella system. Basic properties of the social networks,
including degree distributions, local topological quantities and
degree-degree correlations have been particularly studied in this
paper. The results have proved that peer social networks are small
world networks, as peers are clustered and the path length between
them is small. Moreover, most of the peers (nearly 98.5\%) are pure
resource providers, contributing to the high resource reliability and
availability of P2P networks in resource sharing. Comparatively, free
riding peers that do not contribute any resources are only a small
fraction (less than 1\%) of the whole network. For peers that have
more than one connection, their undirected, directed (including out
and in) and weighted degree distributions follow a clear power-law
distribution. The exponents are greater than 2 for undirected and in
degrees and nearly 1 for out degrees. Investigations on betweenness
and correlations suggest that dynamics of peer social networks are not
dominated by a few highly connected peers. In fact, the peer degrees
are generally disassortative mixing, except some $r_{inin}$ and
$r_{inout}$, suggesting that active providers are connected between
each other and by active requesters.

The collected social networks studied in this paper are only some small
snapshots of the large-scale and continuously changing P2P networks. However,
the kind of study performed here allows us to touch upon the real network
topologies that are difficult to obtain with existing network models. The
analysis results will give useful hints for the future design of effective P2P
systems, by considering their acyclic topologies and small world architecture.
In the future, the joint relation of the social network topology and the
topology of the underlying peer-to-peer network (e.g., Gnutella) will be
studied to examine their commonness and discrepancy.  On top of the kind of
network found in the study, simulations of processes can be enabled to
investigate spreading processes \cite{vespignanibook,mnv04}, modeling of
traffic flow \cite{Echenique2005a} and optimization of network resources
\cite{Echenique2005b}. Based on the current study on peer betweenness and
degree correlations, we will further investigate network hierarchy, peer work
load and dynamic properties of P2P social networks.

\begin{acknowledgments}
  The authors are grateful to Di Liu for his work on Gnutella data collection,
  Dr. Kun Yang and Weibo Chen for their help on
  early data calculation and the anonymous referees for their valuable
  comments. Y.\ M.\ thanks V. Latora for helpful discussions on several aspects
  of this work.  Y.\ M.\ is supported by MEC (Spain) through the Ram\'on y
  Cajal Program and by the Spanish DGICYT project FIS2004-05073-C04-01.
\end{acknowledgments}

\end{document}